\documentclass[]{article}   
\usepackage[]{graphicx}
\usepackage{setspace}
\usepackage[]{graphicx}
\usepackage{amsmath}
\usepackage{amssymb}
\usepackage{subfigure}
\usepackage{epsfig}
\usepackage{wrapfig}
\usepackage{color}
\usepackage{rotating}
\usepackage{float}

\def\##1{{\bf #1}}
\def\=#1{\underline{\underline{#1}}}

\def\r#1{(\ref{#1})}

\def\c#1{\cite{#1}}

\def\les{\left[}
\def\ris{\right]}

\def\ric{\right\}}

\def\.{\mbox{ \tiny{$^\bullet$} }}

\def\eps{\varepsilon}

\def\epso{\eps_{\scriptscriptstyle 0}}
\def\muo{\mu_{\scriptscriptstyle 0}}

\def\etao{\eta_{\scriptscriptstyle 0}}
\def\tauo{\tau_{\scriptscriptstyle 0}}
\def\ko{k_{\scriptscriptstyle 0}}

\def\ro{(\#r,\omega)}
\def\rso{(\#r_{\rm S},\omega)}

\def\Vin{{\cal V}_{\rm in}}
\def\Vout{{\cal V}_{\rm out}}
\def\Surf{{\cal S}}
\def\OUT{_{\rm out}}
\def\IN{_{\rm in}}

\def\un{\hat{\#n}(\#r_{\rm S})}

 \def\ux{\hat{\#{u}}_x}
\def\uy{\hat{\#{u}}_y}
\def\uz{\hat{\#{u}}_z}

\def\deltar{\delta_r}
\def\etar{\eta_r}

\def\sp{\uy}

\def\pinc{\left({-\ux{\tauo}+\uz{\kappa}}\right)/{\ko}}
\def\pref{\left({\ux{\tauo}+\uz{\kappa}}\right)/{\ko}}
\def\ptr{\left({-\ux{\tau}+\uz{\kappa}}\right)/{k}}

\def\as{a_s}
\def\ap{a_p}
\def\rs{r_s}
\def\rp{r_p}
\def\ts{t_s}
\def\tp{t_p}

\begin{document}

\begin{center}

{\sc Classical electromagnetic model of surface states in topological insulators}\\

{Akhlesh Lakhtakia $^{1,\ast}$ and Tom G. Mackay$^{1,2}$}\\

{$^1$Pennsylvania State University, Department of Engineering Science and Mechanics, \\ NanoMM---Nanoengineered Metamaterials Group, 212 EES Building, \\ University Park, PA 16802, USA\\
$^2$University of Edinburgh, School of Mathematics, \\
Edinburgh EH9 3FD, Scotland, United Kingdom\\
$^\ast$ {akhlesh@psu.edu} }

\end{center}


\begin{abstract}
A topological insulator is classically modeled  as an isotropic dielectric-magnetic
with  a magnetoelectric pseudoscalar  $\Psi$ existing in its bulk while its surface is charge-free and current-free. An alternative model is obtained by setting $\Psi\equiv0$ and incorporating surface  charge
 and current
 densities characterized by an admittance $\gamma$.
Analysis of  plane-wave reflection and refraction due to a topological-insulator half space
  reveals that the parameters $\Psi$ and  $\gamma$ arise identically in the reflection and transmission coefficients, implying that the two classical models cannot be distinguished
 on the basis of any scattering scenario. However, as $\Psi$ disappears from   the Maxwell equations   applicable to any region occupied by the topological insulator, and  because surface states exist on topological insulators as protected conducting states,
the alternative model must be chosen.

\end{abstract}


\section{Introduction}
The discovery of topological insulators \cite{Hasan} has prompted  researchers in classical optics \cite{Chang,Liu2013,Liu2014} to examine electromagnetic scattering due to bound objects made of these materials exemplified by chalcogenides
such as Bi$_2$Se$_3$, Bi$_2$Te$_3$, and Sb$_2$Te$_3$.
As a topological insulator is considered to be an isotropic material, its
 frequency-domain constitutive relations are formulated to contain a magnetoelectric pseudoscalar  (denoted
 by $\Psi$ here) in addition to the permittivity scalar $\eps$ and the permeability scalar $\mu$.
 The surface of the topological insulator is assumed to be charge-free and current-free, and the scattering problem can then be solved following textbook techniques \cite{vB}.

Yet, according to condensed-matter theory, surface states exist on topological insulators as protected conducting states \cite{Chang}, and the characteristic electromagnetic responses  of these materials are due to those surface states. Should then a topological insulator's optical response be modeled as due solely to either
\begin{itemize}
\item[I.]
the bulk constitutive parameter $\Psi$ with the surface of the topological insulator being charge-free and current-free,
or
\item[II.]
a surface parameter (denoted by $\gamma$ here) that quantifies the charge density and current density on the surface of the topological insulator?
\end{itemize}
The topological insulator possesses the permittivity $\eps$ and permeability $\mu$ in both models. Both $\Psi$ and $\gamma$ are admittances. Whereas  the magnetoelectric constitutive parameter $\Psi$  mediates between $\#D$ and $\#B$ as well as between $\#H$ and $\#E$ throughout the topological insulator, $\gamma$ is meaningful only on the surface of that material.

This communication is devoted to a comparison of models I and II, through the
fundamental boundary-value problem of reflection and refraction of a plane wave. This problem is described and solved in Sec.~\ref{bvp} for both models. Section~\ref{conc} contains a comparative discussion of the two models. Vectors are underlined. An $\exp(-i\omega{t})$ dependence on time $t$ is implicit, with $\omega$ as the angular frequency and $i=\sqrt{-1}$.

\section{A fundamental boundary-value problem}\label{bvp}
Suppose that all space is divided into two mutually disjoint half spaces $\Vout=\left\{(x,y,z):z<0\right\}$ and $\Vin=\left\{(x,y,z):z>0\right\}$ separated by the surface
$\Surf=\left\{(x,y,z):z=0\right\}$. We need to solve the frequency-domain, macroscopic Maxwell equations
\begin{equation}
\left.\begin{array}{l}
\nabla\.\#B\ro=0\\[4pt]
\nabla\times\#E\ro-i\omega\#B\ro=\#0\\[4pt]
\nabla\.\#D\ro=0\\[4pt]
\nabla\times\#H\ro+i\omega\#D\ro=\#0
\end{array}\ric\,
\label{ME}
\end{equation}
in $\Vout$ and $\Vin$ separately, and impose boundary conditions  on $\Surf$. It is possible to do so for models I and II together.

Let the half space $\Vout$ be vacuous so that the constitutive equations
\begin{equation}
\#D\ro=\epso\#E\ro\,,\qquad
\#H\ro=\muo^{-1}\#B\ro\,,\qquad \#r \in\Vout\,,
\label{conrel0}
\end{equation}
hold, $\epso$ being the permittivity and $\muo$ being the permeability of free space. Equations~\r{ME} can then be written as
\begin{equation}
\left.\begin{array}{l}
\nabla\.\#B\ro=0\\[4pt]
\nabla\times\#E\ro-i\omega\#B\ro=\#0\\[4pt]
\nabla\.\#E\ro=0\\[4pt]
\nabla\times\#B\ro
+i\omega\muo\epso\#E\ro=\#0
\end{array}\ric\,,\quad{\#r\in\Vout}\,
\label{MEout}
\end{equation}
in terms of the primitive field phasors $\#E\ro$ and $\#B\ro$.

The frequency-domain constitutive relations
of the  material occupying $\Vin$ are
\begin{equation}
\left.\begin{array}{l}
\#D\ro=\eps(\omega) \#E\ro +  \Psi(\omega)\#B\ro\\[4pt]
\#H\ro=\mu^{-1}(\omega) \#B\ro  -\Psi(\omega)\#E\ro
\end{array}\ric\,,\qquad  {\#r \in\Vin}\,,
\label{conrel1}
\end{equation}
where $\eps$, $\mu$, and $\Psi$ are functions of $\omega$. Equations~\r{conrel1}
allow us to accommodate model I.
After   substituting Eqs.~\r{conrel1} in Eqs.~\r{ME}
we get
\begin{equation}
\left.\begin{array}{l}
\nabla\.\#B\ro=0\\[4pt]
\nabla\times\#E\ro-i\omega\#B\ro=\#0\\[4pt]
\nabla\. \#E\ro  =0\\[4pt]
\nabla\times \#B\ro  +i\omega\mu(\omega)\eps(\omega)\#E\ro  =\#0
\end{array}\ric\,,\quad{\#r\in\Vin}\,.
\label{MEin2}
\end{equation}
Let us note that $\Psi$ does not appear in the Maxwell equations applied to $\Vin$ after the    convenient but inessential  induction
field phasors $\#D\ro$ and $\#H\ro$ have been translated into the essential primitive field phasors $\#E\ro$ and $\#B\ro$.

When solving an electromagnetic boundary-value problem, it is common to use the  boundary conditions
\begin{equation}
\left.\begin{array}{l}
\un\.\les\#B\OUT\rso-\#B\IN\rso\ris
=0\\[4pt]
\un\times\les\#E\OUT\rso-\#E\IN\rso\ris=\#0\\[4pt]
\un\.\les\#D\OUT\rso-\#D\IN\rso\ris
=\rho_s\rso\\[4pt]
\un\times\les\#H\OUT\rso-\#H\IN\rso\ris=\#J_s\rso
\end{array}\ric\,,\qquad{\#r_{\rm S}}\in\Surf\,,
\label{bc}
\end{equation}
with the unit normal vector $\un$ at $\#r_{\rm S}\in\Surf$ pointing into $\Vout$.
The subscripts ${}_{{\rm in}}$ and ${}_{{\rm out}}$ indicate that the fields in  $\Vin$ and $\Vout$, respectively, are being evaluated on $\Surf$.
The quantities $\rho_s$ and $\#J_s$ are the surface charge density
and the surface current density, respectively.   In order to accommodate model II, we set
\begin{equation}
\left.\begin{array}{l}
\rho_s\rso = \gamma(\omega)\,\un\.\#B\IN\rso\\[4pt]
\#J_s\rso=-\gamma(\omega)\,\un\times\#E\IN\rso
\end{array}\ric\,,\qquad{\#r_{\rm S}}\in\Surf\,,
\label{scdef}
\end{equation}
where $\gamma$  describes the surface states.

Let an arbitrarily polarized plane wave in $\Vout$ be incident on $\Surf$. Then the primitive field phasors
 in $\Vout$ can be written as
 \begin{equation}
\left.\begin{array}{l}
\#E\ro=\Big\{ \les\as\,\sp +\ap \,\pinc\ris \exp(i\tauo{z})
\\[6pt]
\qquad\qquad+\les\rs\,\sp +\rp \,\pref\ris \exp(-i\tauo{z})
\Big\}
\exp(i\kappa {x})
\\[10pt]
\#B\ro=\frac{\ko}{\omega}\Big\{ \les-\ap\,\sp +\as \,\pinc\ris \exp(i\tauo{z})
\\[6pt]
\qquad\qquad+\les-\rp\,\sp +\rs \,\pref\ris \exp(-i\tauo{z})
\Big\}
\exp(i\kappa {x})
\end{array}\right\}
\, , \qquad \#r\in\Vout\, ,
\label{EBincref}
\end{equation}
where $\ko=\omega\sqrt{\muo\epso}$, $\tauo=+\sqrt{\ko^2-\kappa^2}$, and the dependences on $\omega$ are implicit.
Representing the incident plane wave,
the coefficients $\as$ and $\ap$ are presumed to be known. Representing the plane wave
reflected into $\Vout$, the coefficients $\rs$ and $\rp$ are
unknown. Equations~\r{EBincref} satisfy  Eqs.~\r{MEout}.

The primitive field phasors in $\Vin$ are given as
 \begin{equation}
\left.\begin{array}{l}
\#E\ro=  \les\ts\,\sp +\tp \,\ptr\ris \exp(i\tau{z})
\exp(i\kappa {x})
\\[10pt]
\#B\ro=\frac{k}{\omega} \les-\tp\,\sp +\ts \,\ptr\ris \exp(i\tau{z})
\exp(i\kappa {x})
\end{array}\right\}
\, , \qquad \#r\in\Vin
\, ,
\label{EBtr}
\end{equation}
where $k=\omega\sqrt{\mu\eps}$, $\tau = + \sqrt{k^2 - \kappa^2}$,
and
 the coefficients $\ts$ and $\tp$ are unknown.
 Representing the plane wave refracted
into $\Vin$,
these expressions satisfy   Eqs.~\r{MEin2}.

The foregoing expressions were substituted in Eqs.~\r{bc}$_{2,4}$, \r{conrel0}$_{2}$,
\r{conrel1}$_{2}$,
and \r{scdef}$_2$ to determine
$\rs$, $\rp$, $\ts$, and $\tp$ in terms of $\as$ and $\ap$. Thus,
\begin{eqnarray}
\rs&=&
\frac{[(\etar-\deltar)(1+\etar\deltar)-({G}\etao)^2\etar^2\deltar]\as
+2{G}\etao\etar^2\deltar\ap
}{(\etar+\deltar)(1+\etar\deltar)+({G}\etao)^2\etar^2\deltar} \,,
\label{rs-def}
\\
\rp &=&
\frac{[(\etar+\deltar)(1-\etar\deltar)+({G}\etao)^2\etar^2\deltar]\ap+
2{G}\etao\etar^2\deltar\as
}{(\etar+\deltar)(1+\etar\deltar)+({G}\etao)^2\etar^2\deltar} \,,
\label{rp-def}
\\
\ts&=&
\frac{2\etar(1+\etar\deltar)\as+2{G}\etao\etar^2\deltar\ap}{(\etar+\deltar)(1+\etar\deltar)+({G}\etao)^2\etar^2\deltar}\,,
\label{ts-def}
\\
\tp&=&
\frac{2\etar(\etar+\deltar)\ap-2{G}\etao\etar^2\as}{(\etar+\deltar)(1+\etar\deltar)+({G}\etao)^2\etar^2\deltar} \,,
\label{tp-def}
\end{eqnarray}
where
\begin{equation}
\label{defG}
G=\Psi+\gamma\,,\qquad
\etao=\sqrt{\frac{\muo}{\epso}}\,,\qquad
\deltar=\frac{\tau/k}{\tauo/\ko}\,,\qquad
\etar=\sqrt{\frac{\epso\mu}{\eps\muo}}\,.
\end{equation}

We have verified that Eqs.~\r{rs-def}--\r{tp-def} satisfy Eqs.~\r{bc}$_{1,3}$ and \r{scdef}$_1$.
Moreover, Eqs.~\r{rs-def}--\r{tp-def} simplify to the standard results \c{Lakh2,Iskander}
\begin{equation}
\left.\begin{array}{ll}
\rs =  \displaystyle{\as
\frac{\etar-\deltar}{\etar+\deltar}}
\,,\qquad
&\rp= \displaystyle{\ap
\frac{1-\etar\deltar}{1+\etar\deltar}}
\\[8pt]
\ts= \displaystyle{\as
\frac{2\etar}{\etar+\deltar}}
\,,\qquad
&\tp= \displaystyle{\ap
\frac{2\etar}{1+\etar\deltar}}
\end{array}\ric\,
\end{equation}
for $\Psi=\gamma=0$.

\section{Discussion and Conclusion}\label{conc}

Equations~\r{rs-def}--\r{tp-def} can be recast in matrix form as
\begin{equation}
\left[\begin{array}{c} \rs \\ \rp\end{array}\right]
=
\left[\begin{array}{cc} r_{ss} & r_{sp} \\ r_{ps} & r_{pp}\end{array}\right]
\left[\begin{array}{c} \as \\ \ap\end{array}\right]\,,
\qquad
\left[\begin{array}{c} \ts \\ \tp\end{array}\right]
=
\left[\begin{array}{cc} t_{ss} & t_{sp} \\ t_{ps} & t_{pp}\end{array}\right]
\left[\begin{array}{c} \as \\ \ap\end{array}\right]\,.
\end{equation}
The elements of the 2$\times$2 matrixes  have either both subscripts identical or two
 different subscripts. The elements with both subscripts identical indicate co-polarized
 reflection or refraction,
 the remaining elements indicating cross polarization. Both cross-polarized reflection and refraction in Eqs.~\r{rs-def}--\r{tp-def} are due to $G$.

Equations~\r{rs-def}--\r{tp-def} do not contain $\Psi$ and $\gamma$ separately, but their sum $G$ instead. Thus, measurements of the reflection coefficients $\rs$ and $\rp$
(or the transmission coefficients $\ts$ and $\tp$, if at all possible)
 cannot be used to discriminate between models I ($\gamma=0$) and II ($\Psi=0$). Equations~\r{bc}$_{4}$ and \r{scdef}$_2$ together make it clear that measurements of the reflection and transmission coefficients of
a slab made of a topological insulator  cannot be used to discriminate between the two models; not only that, the solution of every scattering problem will depend on $G$, not on $\Psi$ alone or $\gamma$ alone.

This impasse can be resolved on realizing that surface states exist on topological insulators as protected conducting states, and the characteristic behavior of these materials is due to those surface states. Furthermore, $\Psi$ vanishes from the Maxwell equations \r{MEin2}
applicable to $\Vin$ occupied by the topological insulator; indeed, $\Psi$ would vanish even if the topological insulator were bianisotropic \cite{LM2015}.
For both of these reasons, we must choose model II, which also satisfies the Post constraint $\Psi\equiv0$ \cite{Post}.

As the material occupying   $\Vin$ is isotropic and achiral,
cross-polarized reflection in this
problem has been taken to arise from the Lorentz nonreciprocity inherent
in Eqs.~\r{conrel1}  \cite{Krowne}.
 But now we see that surface states described by Eqs.~\r{scdef} by themselves
are capable of yielding cross-polarized reflection, which is therefore not an indication
on Lorentz noneciprocity.

\vskip3mm

\noindent{\bf Acknowledgments.}
AL is grateful to the Charles Godfrey
Binder Endowment at Penn State for ongoing support of his research.
TGM acknowledges the support of EPSRC grant EP/M018075/1.

\end{document}